\documentclass{article}
\usepackage{graphicx}

\begin{document}
\bibliographystyle{unsrt}

\title{Possible Mechanism for Superconductivity in Sulfur---Common
Theme for Unconventional Superconductors?} 
\author{Eric Lewin Altschuler\footnote{School of Medicine,
University of California, San Diego, 9500 Gilman Drive, 0606,
La Jolla, CA 92093-0606, email: elaltsch\@sdcc3.ucsd.edu} and
Martin Lades\footnote{Livermore, CA, 94550}}
\date{\today}
\maketitle
\begin{abstract}
Sulfur has recently been found to be a superconductor at high
pressure.  At $\sim$93~GPa $T_c$ is 10.1~K, and the sulfur is in a
base-centered orthorhombic (b.c.o.) structure.  At $\sim$160~GPa $T_c$
is 17~K and sulfur is in a rhombohedral ($\beta$-Po) structure.  The
mechanism for superconductivity in sulfur is not known; in particular,
a band-structure calculation does not find superconductivity in sulfur
until 500~GPa.  Following from work by Anderson, in a 2D strongly
interacting, non-fermi liquid system with some degree of disorder at
$T=0$, the only known conducting state is a superconductor.  Following
this idea it has been suggested that both the $HT_c$ cuprates and 2D
electron gas systems are superconductors with planar conducting
planes.  Similarly, here we suggest that the mechanism for
conductivity in sulfur are 2D conducting planes which emerge as the
planar rings in sulfur at low pressure pucker at higher pressures
(b.c.o. and $\beta$-Po).  As well, we note some other consequences for
study of $HT_c$ materials of Anderson's work.
\end{abstract}

\begin{figure}[htb]
\begin{center}
\includegraphics[width=1.5in]{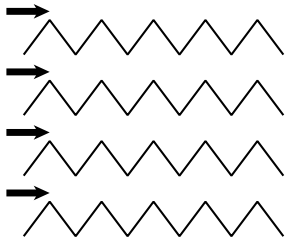}
\caption{\small \leftskip=2pc \rightskip=2pc Emergent potential
conducting planes in sulfur at high pressure.  In this highly
schematic figure vertices represent sulfur atoms and lines bonds
between the atoms.  There are more rows of atoms both in the plane of
the paper and in a plane perpendicular to the paper going into the
paper.  The direction of conduction is indicated by the arrow and the
proposed 2D conduction plane is perpendicularly going into the paper.
Notice that the proposed conduction plane emerges due to the
puckering.}
\label{img:fig1}
\end{center}
\end{figure}

Recently Struzhkin {\it et al.}~\cite{struzhkin} have found that at
high pressures sulfur becomes a superconductor.  At low pressure
sulfur is an insulator with a planar ring structure.  Struzhkin {\it
et al.} find that at $\sim$93~GPa sulfur is a superconductor with
$T_c$ of 10.1 K.  At this pressure sulfur adopts a base-centered
orthorhombic (b.c.o.) structure~\cite{akaham} in which the planar
rings are now puckered.  At $\sim$160~GPa Struzhkin {\it et al.} find
$T_c$ of 17 K.  At this pressure sulfur is in a rhombohedral phase
($\beta$-Po structure)~\cite{luo} which also features puckered rings.
The mechanism for superconductivity of sulfur is not completely well
understood.  Indeed, Struzhkin {\it et al.} note that using
band-structure calculations of electron-phonon interactions Zakharov
and Cohen~\cite{zakharov} found sulfur to be superconducting above 550
GPa, but not at the much lower pressure in which superconductivity was
found experimentally.  Here we suggest that similarly to proposed
mechanisms of superconductivity in copper oxide materials, and 2D
electron gases at low temperatures~\cite{phillips}, (which mechanisms
we grant themselves are controversial), is due to conduction in 2D
planes, which, in the case of sulfur emerge in puckered rings
Fig.~\ref{img:fig1}.

Philips {\it et al.}~\cite{phillips} seizing upon recent experimental
observations~\cite{kravchenko,simmons} finding that in a number of
systems 2D electron gases at low temperature are conductors, in
contradiction to theory which predicted the electron gas to be an
insulator, suggested that not only is the electron gas a conductor,
but a superconductor. They provide a number of arguments to support
this notion including the features of the transition from insulator to
conductor in the electron gas system being reminiscent of an
insulator-superconducting transition; there exists a critical magnetic
field above which conductivity is destroyed; and the
insulating-conducting transition is near an electron crystal state in
which large charge retardation effects could possibly lead to Cooper
pairing.  Furthermore, with reference to a classic paper written by
Anderson~\cite{anderson}, they note that in 2D at T=0 the only known
conducting non-Fermi liquid state in the presence of disorder with
zero magnetic field, is a superconductor.  In general Anderson's
paper~\cite{anderson} emphasizes that often in the presence of some
disorder, a superconducting state can be more stable and less likely
to be abolished than other conducting states. High pressures may
configure sulfur into such a strongly interacting non-Fermi liquid
state with emergent planes. A similar mechanism may explain
superconductivity in oxygen at high
pressure~\cite{shimizu}. Experiments consistent with this idea would
include the finding of conduction preferentially in the direction
indicated in the figure.

The application of the idea of Anderson's paper to unconventional
superconducting materials has a number of current and future
applications:
(1) It has recently been reported that $T_c$ can nearly be doubled in
certain superconducting perovskites by making thin films of the
material under epitaxial strain~\cite{locquet}.  This result may be
hard to explain by theories proposing a single mechanism responsible
for pairing in the superconducting state.  However, assuming, at least
{\it in arguendo} that 2D planes are important for superconductivity
in these materials, this result is much easier to understand from the
perspective of Anderson's paper: Perhaps there are a number of
different contributions to the pairing mechanism, but regardless of
the nature or number of such contributions, $T_c$ in a conducting (and
thus superconducting) material will rise proportionally to a reduction
in the localizing ability of the host state which is accomplished by
growth of a material under epitaxial strain.
(2) Theoretically and computationally it might be useful to try to
find strongly interacting non-Fermi liquid systems which are not
superconductors in the presence of some disorder, to help steer
experiments from non-productive paths.  As well, it would be helpful
to try to find other general classes of geometries or materials which
are strongly interacting non-Fermi liquids and possible conductors.
(3) Experimentalists should appreciate that, especially when studying
systems which are predicted to be insulators, if a material is a
conductor, it might be a superconductor.  As well, it might be
possible, for example in 2D systems, to screen large numbers of
materials looking for conductors.

\bibliography{sulscon}
\end{document}